# Design Optimal Backstepping Controller for Quadrotor Based on Lyapunov Theory for Disturbances Environments

Dong LT Tran[1], Thanh C Vo[1], Hoang T Tran[1], Minh T Nguyen[2,3,*], Hai T. Do[2]

[1] Center of Electrical Engineering, Duy Tan University, Da Nang 550000, Vietnam

[2] Thai Nguyen University of Technology (TNUT), Thai Nguyen 240000, Viet Nam

[3] Thai Nguyen University, Thai Nguyen 240000, Viet Nam

*Email: nguyentuanminh@tnut.edu.vn

## Abstract

Various control methods have been studied to control the position and attitude of quadrotors. There are some differences in the mathematical equations between the two types of quadrotor configurations that lead to different control efficiency in disturbance environments. This paper described the nonlinear backstepping approach based on the Lyapunov function theory and Krasovaskii-LaSalle Principle for the quadrotor's control system, which can provide the stability of all system states during the tracking of the desired trajectory. Accordingly, a mathematical model of the cross quadrotor configuration together with the controller has been built to stabilize the altitude and position of the quadrotor. To clarify the effectiveness of this method with the selected quadrotor configuration, we compare it with a traditional PID controller in an environment affected by disturbances. The simulation results in Matlab show satisfactory stability of the quadrotor flight and following certain trajectories, confirming the accuracy and validity of the control method.

## 1. Introduction

There are two basic types of quadrotor configurations: a plus configuration and a cross configuration. In general, the mathematical model of the two configurations above is similar. Accordingly, many quadrotor control methods have been studied, designed and compared to confirm their effectiveness. There have been many previous studies that made a comparison between PID and Backstepping methods but based on plus quadrotor configuration. For example, the result in [1, 2] show that the backstepping controller is better than the PID controller for noise attenuation. When dealing with wind disturbance, the Backstepping controller shows better disturbance rejection than the PID controller. Meanwhile, many studies show that the cross con-figuration provides more stable control than the plus configuration [3, 4]. In greater detail, the distance away from the axis of rotation dictates the torque created by the motors. Consider a quadrotor with each arm l distant from the quadrotor's center of mass. As mentioned in our previous study [5], with the cross configuration, thrust is applied at a distance of   (or  ) since the arms are at a 45-degree angle to the axis of rotation, which produces more rotational acceleration than plus configuration. Accordingly, the cross quadcopter has been proven to be more stable in maneuvering due to the absence of residual rotational velocity [6]. Given the numerous benefits associated with the implementation of Backstepping approach along with the optimization of control problems of the cross quadrotor configuration com-pared to the plus quadrotor configuration, in this paper, we describe in detail the mathematical modelling of quadrotors and propose a nonlinear backstepping meth-od on the Lyapunov function and Krasovaskii-LaSalle Principle for the quadrotor, which provides the ability to track the desired trajectory asymptotically stable under disturbance environments. Our main contribution is to demonstrate the asymptotic stability of the cross quadrotor configuration based on the equation of state already in our previously performed studies in [5].

## 2. Equation of Motion Quadrotor

In our previous work [5, 7], we have described the design and the mathematical equations describing the kinematics of the quadrotor. Thus, from equation 11 in [5], the kinematic model of the cross quadrotor is as follows:

$$\begin{cases} \dot{x} = w[s(\varphi)s(\psi) + c(\varphi)c(\psi)s(\theta)] - v[c(\varphi)s(\psi) - c(\psi)s(\varphi)s(\theta)] \\ \quad + u[c(\psi)c(\theta)] \\ \dot{y} = v[c(\varphi)c(\psi) + s(\varphi)s(\psi)s(\theta)] - w[c(\psi)s(\varphi) - c(\varphi)s(\psi)s(\theta)] \\ \quad + u[c(\theta)s(\psi)] \\ \dot{z} = w[c(\varphi)c(\theta)] - u[s(\theta)] + v[c(\theta)s(\varphi)] \\ \dot{\varphi} = p + r[c(\varphi)t(\theta)] + q[s(\varphi)t(\theta)] \\ \dot{\theta} = q[c(\varphi)] - r[s(\varphi)] \\ \dot{\psi} = r\dfrac{c(\varphi)}{c(\theta)} + q\dfrac{s(\varphi)}{c(\theta)} \end{cases} \quad (1)$$

where $\begin{bmatrix} x & y & z & \varphi & \theta & \psi \end{bmatrix}^T$ is the vector containing the quadrotor's linear and angular position in the earth frame. The dynamic model of quadrotor is obtained from Newton–Euler approach. Both linear and angular dynamics are addressed in equation 2 and equation 3, respectively.

$$m\ddot{r} = \begin{bmatrix} 0 \\ 0 \\ -mg \end{bmatrix} + R \begin{bmatrix} 0 \\ 0 \\ u_1 \end{bmatrix} \quad (2)$$

$$I \begin{bmatrix} p \\ q \\ r \end{bmatrix} = u_2 - \begin{bmatrix} p \\ q \\ r \end{bmatrix} \times I \begin{bmatrix} p \\ q \\ r \end{bmatrix} \quad (3)$$

where $[p \; q \; r]^T$ is the body angular accelerations measured by the gyroscope; $m$ is the system mass; $I$ is the system moment of inertia; $u_1$ is the thrust input and $u_2$ is the moment input ($3 \times 1$ vector) and the rotation matrix R is as follows:

$$R = \begin{pmatrix} c(\theta)c(\psi) & s(\varphi)s(\theta)c(\psi) - c(\varphi)c(\psi) & c(\varphi)s(\theta)c(\psi) + s(\varphi)s(\psi) \\ c(\theta)s(\psi) & c(\varphi)c(\psi) + s(\varphi)s(\theta)s(\psi) & c(\varphi)s(\theta)s(\psi) - s(\varphi)s(\psi) \\ -s(\theta) & s(\varphi)c(\theta) & c(\varphi)c(\theta) \end{pmatrix} \quad (4)$$

To synthesize adaptive laws, it is assumed that some uncertainty remains with respect to the main coefficients related to the aerodynamic torques, inertia matrix and mass of the system. The main control objective is here the stabilization of bank and pitch angles while tracking heading and altitude trajectories. Following the classic work in [8, 9], Lyapunov stability theory has served as acornerstone of most control stability work and control law development tasks for nonlinear systems. Applying this theory for the UAV stabilization and control [10, 11]. A good controller should be able to reach a desired position and a desired yaw angle while guaranteeing stability of the pitch and roll angles. Hence, the mathematical model (1) can be used to write the dynamic system in state-space form $\dot{X} = F(X,U)$, with the vector of control inputs $U = [u_1, u_2, u_3, u_4]^T$ and following state vector $X = [x_1, x_2, x_3, x_4, x_5, x_6, x_7, x_8, x_9, x_{10}, x_{11}, x_{12}]^T$ mapped to the degrees of freedom of the quadrotor in the following manner, as $X = [\varphi, \dot{\varphi}, \theta, \dot{\theta}, \psi, \dot{\psi}, x, \dot{x}, y, \dot{y}, z, \dot{z}]^T$. The state vector defines the position of the quadrotor in space and its angular andlinear velocities. By simply choosing:

$$\begin{cases} a_1 = \left( \frac{I_{yy} - I_{xx}}{I_{xx}} \right); a_2 = \left( \frac{I_{zz} - I_{xx}}{I_{yy}} \right); a_3 = \left( \frac{I_{xx} - I_{yy}}{I_{zz}} \right) \\ b_1 = \frac{1}{I_{xx}}; b_2 = \frac{1}{I_{yy}}; b_3 = \frac{1}{I_{zz}} \end{cases} \quad 5)$$

Using these state variables and the parameters, the dynamic model can be written as:

$$\begin{bmatrix} \dot{x}_1 \\ \dot{x}_2 \\ \dot{x}_3 \\ \dot{x}_4 \\ \dot{x}_5 \\ \dot{x}_6 \\ \dot{x}_7 \\ \dot{x}_8 \\ \dot{x}_9 \\ \dot{x}_{10} \\ \dot{x}_{11} \\ \dot{x}_{12} \end{bmatrix} = \begin{bmatrix} x_2 \\ a_1 x_4 x_6 + b_1 u_2 \\ x_4 \\ a_2 x_2 x_6 + b_2 u_3 \\ x_6 \\ a_3 x_2 x_4 + b_3 u_4 \\ x_8 \\ -g + b_4 [(c(x_1)c(x_3)]u_1 \\ x_{10} \\ b_4 [c(x_5)s(x_3) + s(x_5)c(x_3)s(x_1)]u_1 \\ x_{12} \\ b_4 [s(x_5)a(x_3) - c(x_5)c(x_3)s(x_1)]u_1 \end{bmatrix} \quad (6)$$

## 3. Optimal Backstepping Controller based on Lyapunov

Using the backstepping approach as a recursive algorithm for the control laws synthesis according to the high-order nonholonomic constraints, we simplify all the stages of calculation concerning the tracking errors and Lyapunov functions. The proposed backstepping control block diagram for control of a cross quadrotor is show in figure. 1.

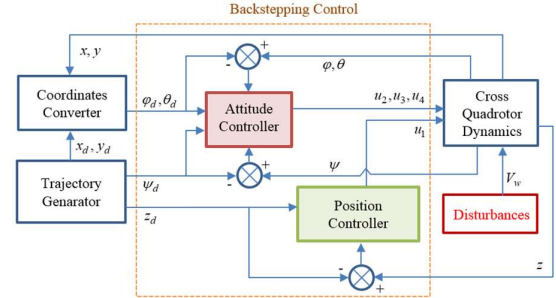

**Fig. 1.** Block Diagram of Backstepping Control

From equation 6, we have the states $[x_1 \; x_3 \; x_5 \; x_7]$ are the roll $(\varphi)$, pitch $(\theta)$, yaw $(\psi)$ and altitude $(z)$; The states $[x_2 \; x_4 \; x_6 \; x_8]$ are in order the rates of change of $(\varphi, \; \theta, \; \psi, \; z)$. Therefore, we divided into four subsystems are as follow:

$$\begin{bmatrix} \dot{x}_1 \\ \dot{x}_2 \end{bmatrix} = \begin{bmatrix} x_2 \\ a_1 x_4 x_6 + b_1 u_2 \end{bmatrix} \quad \varphi \text{ equation} \quad (7)$$

$$\begin{bmatrix} \dot{x}_3 \\ \dot{x}_4 \end{bmatrix} = \begin{bmatrix} x_4 \\ a_2 x_2 x_6 + b_2 u_3 \end{bmatrix} \quad \theta \text{ equation} \quad (8)$$

$$\begin{bmatrix} \dot{x}_5 \\ \dot{x}_6 \end{bmatrix} = \begin{bmatrix} x_6 \\ a_3 x_2 x_4 + b_3 u_4 \end{bmatrix} \quad \psi \text{ equation} \quad (9)$$

$$\begin{bmatrix} \dot{x}_7 \\ \dot{x}_8 \end{bmatrix} = \begin{bmatrix} x_8 \\ -g + b_4 [(c(x_1)c(x_3)]u_1 \end{bmatrix} \quad z \text{ equation} \quad (10)$$

The $(\varphi \; \theta \; \psi)$ angle and $(z)$ equation is strictly in a feedback form, which means that only the second equation in each of the above substem is affected by an input. With $i = 1,3,5,7$ the following simple positive definite Lyapunov Function is chosen:

$$V_i = \frac{1}{2} e_i^2 \quad (11)$$

where: $e_{1,3,5}$ is the error between the desired and the actual $(\varphi \ \theta \ \psi)$ angle; $e_7$ is the error between the desired and the actual $(z)$ and value of $e_i$ is defined by the equation:

$$e_i = x_{ir} - x_i \quad (12)$$

The time derivate of the function defined in equation 11 is:

$$\dot{V}_i = e_i \dot{e}_i = e_i(\dot{x}_{ir} - \dot{x}_i) = e_i(\dot{x}_{ir} - x_{i+1}) \quad (13)$$

The system is guaranteed to be stable if the time derivative of the positive denite Lyapunov function is negative semidefinite, according to the Krasovaskii-LaSalle principle [12]. As shown in equation 14, a positive definite bounding function that is a bound on $V_i$ is chosen.

$$\dot{V}_i = e_i(\dot{x}_{ir} - x_{i+1}) \leq -c_i e_i^2 \quad (14)$$

here $c_i$ is a positive constant ($V_i$ has a limit of $c > 0$ when $time \to \infty$) [12]. In order to meet inequality 14, the virtual control input is set to:

$$(x_{i+1})_{desired} = \dot{x}_{ir} + c_i e_i \quad (15)$$

A deviation of the state $x_{i+1}$ from its desired value is denoted by a new error variable $e_{i+1}$

$$e_{i+1} = x_{i+1} - \dot{x}_{ir} - c_i e_i \quad (16)$$

According to equations $13 \div 16$, $\dot{V}_i$ can be rewritten as following:

$$\dot{V}_i = e_i \dot{e}_i = e_i(\dot{x}_{id} - x_{i+1})$$
$$= e_i(\dot{x}_{id} - (e_{i+1} + \dot{x}_{id} + c_i e_i)) \quad (17)$$
$$= -e_i e_{i+1} - c_i e_i^2$$

To obtain a positive definite $V_{i+1}$, augment the first Lyapunov function $V_i$ by a quadratic term in the second variable $e_{i+1}$.

$$V_{i+1} = V_i + \frac{1}{2} e_{i+1}^2 = -e_i e_{i+1} - c_i e_i^2 + e_{i+1}(\dot{x}_{i+1} - \ddot{x}_{ir} - c_i \dot{e}_i) \quad (18)$$

Choosing a positive definite bounding function, substituing $i = 1,3,5,7$ and the model $\dot{x}_{i+1}$ leads to the following:

$$\begin{cases} -e_1 e_2 - c_1 e_1^2 + e_2(a_1 x_4 x_6 + b_1 u_2 - \ddot{x}_{1d} - c_1 \dot{e}_1) \leq -c_1 e_1^2 - c_2 e_2^2 \\ -e_3 e_4 - c_3 e_3^2 + e_4(a_2 x_2 x_6 + b_2 u_3 - \ddot{x}_{3d} - c_3 \dot{e}_3) \leq -c_3 e_3^2 - c_4 e_4^2 \\ -e_5 e_6 - c_5 e_5^2 + e_6(a_3 x_2 x_4 + b_3 u_4 - \ddot{x}_{5d} - c_5 \dot{e}_5) \leq -c_5 e_5^2 - c_6 e_6^2 \\ -e_7 e_8 - c_7 e_7^2 + e_8(-g + b_4 [c(x_1)c(x_3)] u_1 - \ddot{x}_{7d} - c_7 \dot{e}_7) \leq -c_7 e_7^2 - c_8 e_8^2 \end{cases} \quad (19)$$

Using the equality case of equation 19, we drawable:

$$\begin{cases} u_2 = \frac{1}{b_1}(\ddot{x}_{1d} + c_1 \dot{e}_1 - a_1 x_4 x_6 + e_1 - c_2 e_2) \\ u_3 = \frac{1}{b_2}(\ddot{x}_{3d} + c_3 \dot{e}_3 - a_2 x_2 x_6 + e_3 - c_5 e_5) \\ u_4 = \frac{1}{b_3}(\ddot{x}_{5d} + c_5 \dot{e}_5 - a_3 x_2 x_4 + e_5 - c_6 e_6) \\ u_1 = \frac{1}{b_4 c(x_1) c(x_3)}(\ddot{x}_{7d} + c_7 \dot{e}_7 - c_8 e_8 + g + z_7) \end{cases} \quad (20)$$

## 4. Simulation Results

Next, the installation parameters are selected to approximate the real model - platform F450 according to document [13]. Details of these parameters are shown in Table 1:

**Table 1.** Specifications of the quadrotor model

| Parameters | Value | Unit | Desc |
|---|---|---|---|
| g | 9.81 | m/s² | Gravitational acceleration |
| l | 0.225 | m | Distance from quadrotor center to rotor center |
| m | 2 | kg | Quadrotor mass |
| $I_{xx}$ | 0.0035 | kg/m² | Moment of inertia of the frame along the x axis |
| $I_{yy}$ | 0.0035 | kg/m² | Moment of inertia of the frame along the y axis |
| $I_{zz}$ | 0.0050 | kg/m² | Moment of inertia of the frame along the z axis |

With the available parameters, ignoring coriolis force or nonlinear aerodynamic phenomena, the quadrotor model is designed as shown in figure 2. Accordingly, a spiral trajectory has been proposed for this work's performance analysis.

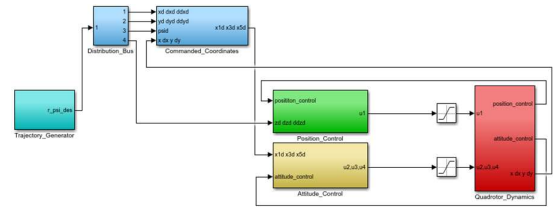

**Fig. 2.** Simulink Model of a Quadrotor System

The research compares the simulation results of the backstepping control approach and traditional PID control. In the test case, we add to the system the disturbance factor which is the wind of velocity. This disturbance factor is a ladder function signal with an initial value is 0, the final value is 6.0, a sampling time is 0, and the interval of output to the final value since the start of the simulation is 25. We can easily set by Step block in Simulink. The positional and attitude are no longer the same. Simulation results between the two methods are shown in figure. 3 and figure 4.

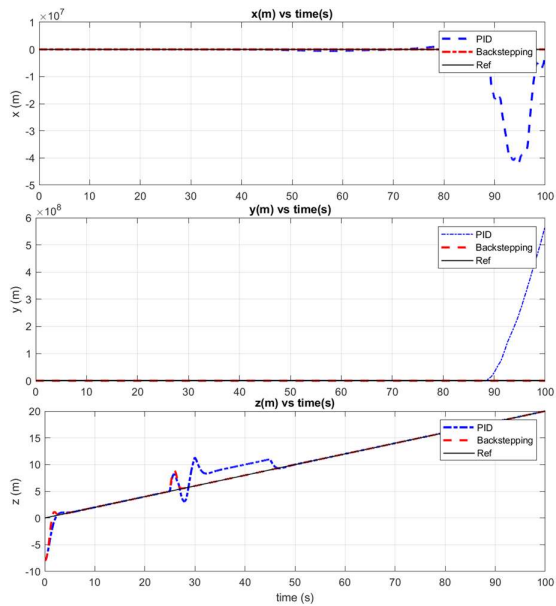

**Fig. 3.** Position and Attitude vs Time of quadrotor under disturbance at 25s

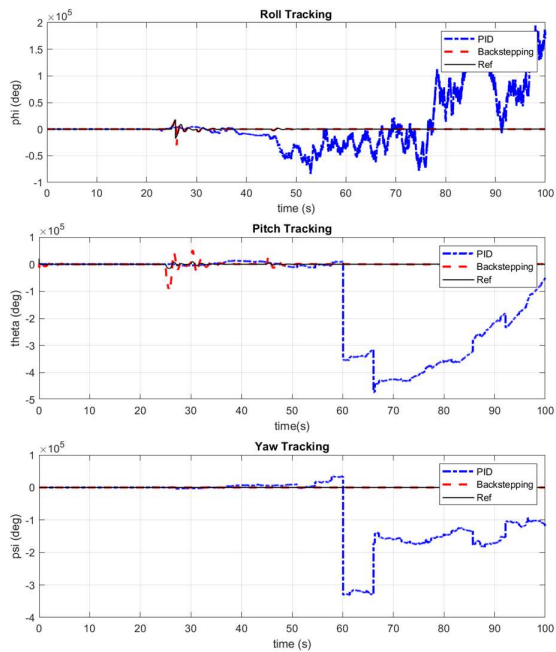

**Fig. 4.** Orientation vs Time of quadrotor under disturbance at 25s

The results show the response of the positions, orientations, control inputs and trajectory error in the presence of disturbances. It can be concluded from the plots that the PID control does not perform as well as Backstepping. At the time=25s, control error occurs in both controllers. The initial errors of the Backstepping controller seem to be larger than that of the PID controller. However, after only about 5 - 7 seconds, the UAV returned to equilibrium with asymptotic deviation to zero for the Backstepping controller. In contrast to the PID controller, the deviations are getting larger and the time to return to the old state is longer until the end of the simulation. The simulations reveal that, while the backstepping controller does not become unstable when perturbations are added, it does exhibit significant steady-state inaccuracy. The solution could be to use integral Backstepping. In short, the Lyapunov function approach in the Backstepping controller assures stability for linear state space models while being more efficient than a PID controller. The simulations reveal that, while the backstepping controller does not become unstable when perturbations are added, it does exhibit significant steady-state inaccuracy. The solution could be to use integral backstepping. In short, the Lyapunov function approach in the backstepping controller assures stability for linear state space models while being more efficient than a PID controller.

## 5. Conclusions

This work presents the implementation of a backstepping control approach for achieving dynamic control of a cross quadrotor configuration. Accordingly, a control strategy utilising the Lyapunov function and Krasovaskii-LaSalle Principle was devised to govern the position and attitude of the quadrotor subsystem. The performance of the backstepping method was evaluated using a disturbances scenario featuring a wind velocity of 6 m/s. The simulation results demonstrated the effectiveness of the suggested control approach. The next research integrated Backstepping and PID methods in a self-switching system based on machine learning methods to create a more flexible and optimal control system for the quadrotor.

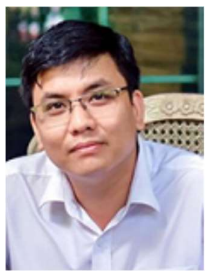
**PhD Student. Dong LT. Tran** received the B.S. degree from the Duy Tan University, Da Nang, Vietnam, in 2009, the M.S. degree from the Duy Tan University, Da Nang, Vietnam, in 2012, and became a PhD student at Thai Nguyen University from 2022 to the present. He is currently the Vice Director of The Center of Electrical Engineering, Duy Tan University, Da Nang, Vietnam, and the Lecture of the Faculty of Electronic Engineering, Duy Tan University, Da Nang, Vietnam. He can be contacted at email: tranthangdong@duytan.edu.vn.

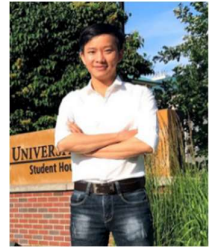
**Eng. Thanh C. Vo** received a degree in Mechatronic Engineering from Danang University of Technology, Vietnam in 2013. Engineer Thanh is currently an expert at Center for Electrical and Electronic Engineering, Duy Tan University, Vietnam (CEE). He has interest and expertise in research topics in the field of embedded programming, automatic control. He can be contacted at email: vochithanh.cdt@gmail.com.

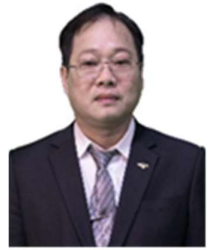
**Dr. Hoang T. Tran** received a master's degree in network and electrical systems in 2009 from the University of Danang; Defending the PhD thesis majoring in Electronics and Communication Technology in 2016 at the University of Technology, Vietnam National University, Hanoi. A guide to synthesizing data sensors, wireless network transformation and the fields of control automation. He is currently the Director of The Center of Electrical Engineering, Duy Tan University, Da Nang, Vietnam, and the Lecture of the Faculty of Electronic Engineering, Duy Tan University, Da Nang, Vietnam. He can be contacted at email: tranthuanhoang@duytan.edu.vn.

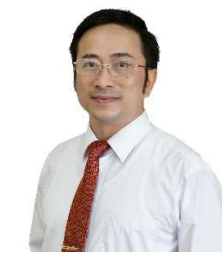
**Assoc. Prof. Dr. Minh T. Nguyen** is currently the director of international training and cooperation center at Thai Nguyen University of Technology, Vietnam, and also the director of advanced wireless communication networks (AWCN) lab. He has interest and expertise in a variety of research topics in the communications, networking, and signal processing areas, especially compressive sensing, and wireless/mobile sensor networks. He serves as technical reviewers for several prestigious journals and international conferences. He also serves as an editor for wireless communication and mobile computing journal and an editor in chief for ICSES transactions on computer networks and communications. He is in the 2023 Stanford's list World Top 2% scientists. He can be contacted at email: nguyentuanminh@tnut.edu.vn.